# A Comprehensive Survey on the Multiple Travelling Salesman Problem: Applications, Approaches and Taxonomy

Omar Cheikhrouhou[a,*], Ines Khoufi[b]

[a] *College of CIT, Taif University, P.O. Box 11099, Taif 21944, Saudi Arabia.*
[b] *SAMOVAR, Télécom SudParis, Institut Polytechnique de Paris, France.*
*CES Laboratory, University of Sfax, Tunisia.*

**Abstract**

The Multiple Travelling Salesman Problem (MTSP) is among the most interesting combinatorial optimization problems because it is widely adopted in real-life applications, including robotics, transportation, networking, etc. Although the importance of this optimization problem, there is no survey dedicated to reviewing recent MTSP contributions. In this paper, we aim to fill this gap by providing a comprehensive review of existing studies on MTSP. In this survey, we focus on MTSP's recent contributions to both classical vehicles/robots and unmanned aerial vehicles. We highlight the approaches applied to solve the MTSP as well as its application domains. We analyze the MTSP variants and propose a taxonomy and a classification of recent studies.

*Corresponding author
  *Email addresses:* `o.cheikhrouhou@tu.edu.sa` (Omar Cheikhrouhou),
`ines.khoufi@telecom-sudparis.eu` (Ines Khoufi)

Contents

1. Introduction and Motivation

The Multiple Traveling Salesman Problem (MTSP) is a generalization of the well-known Traveling Salesman Problem (TSP), where multiple salesmen are involved to visit a given number of cities exactly once and return to the initial position with the minimum traveling cost. MTSP is highly related to other optimization problems such as Vehicle Routing Problem (VRP) [1] and Task Assignment problem [2]. Indeed, MTSP is a relaxation of VRP with neither considering the vehicle capacity or customer demands. MTSP also shares some characteristics with the task assignment problem, however, MTSP does not allow multiple visits and sub tours. Thus, a solution to MTSP can be used to address VRP or Task Assignment optimization problem.

MTSP is one of the most important optimization problems, and it has been applied in several real-life scenarios long ago. Depending on the application requirements, the salesmen in MTSP can be represented by ground vehicles such as robots or trucks, or by flying vehicles such as Unmanned Aerial Vehicles (UAVs) known also as drones. Whereas the cities to be visited by the salesmen can have different representations, such as customers in transportation and delivery services, sensor nodes for Wireless Sensor Networks data collection, targets in military applications, victims in emergency missions and critical sites in disaster management applications.

Considering the importance of such optimization problem, we dedicated this survey to review, analyse and discuss recent contributions on MTSP while highlighting the application domains and the approaches applied to solve the problem for both classical vehicles (i.e. ground robots, vehicles, trucks) and flying vehicles (i.e. UAVs and drones).

In the context of classical vehicles, several surveys on vehicle routing problem [3, 4, 5, 6] and traveling salesman problem [7, 8] have been proposed in the literature to review the different variants of those optimization problems as well as the approaches applied to solve them. To name a few, the authors in [7] presented a survey on routing problems for robotic systems. This survey discussed some routing solutions based on TSP and VRP and their extension to satisfy the robotic system's constraints. These solutions include the Dubins TSP (DTSP), a generalization of the TSP in which a path is composed of Dubins curves; TSP with Neighborhoods (TSPN), where a neighborhood is associated with each city and the salesman needs to visit any neighborhood; the Multiple Traveling Salesman Problem (MTSP) and finally the Generalized Asymmetric TSP (GATSP).

The study in [8] introduced a comparative analysis of evolutionary algorithms for the Multi-Objective Travelling Salesman Problem (MOTSP). The authors focused on only five evolutionary algorithms, namely NSGA-II, NSGA-III, SPEA-2, MOEA/D and VEGA, in order to determine the most suitable algorithm for the MOTSP problem.

Although it was cited in several studies, MTSP was only mentioned in some sections but was not the subject of any of the above surveys. Indeed, the only survey on MTSP proposed by Bektas [9] dates back to 2006, and which reviewed MTSP and its applications, highlighted some formulations, and described some exact and heuristic solutions.

Therefore, there is a lack of analysis studies devoted to MTSP. To fill this gap, the present survey is dedicated to review recent contributions that focus on either MTSP and its variants or on the use of MTSP to solve real-life problems.

In the context of flying vehicles, the study in [10] presented a survey focusing on UAV optimization problems and proposed different variants of TSP and VRP for UAVs. Another interesting study is introduced in [11], where the authors reviewed contributions on UAV trajectory optimization and UAV routing. They also provided a formal definition of the UAV Routing and Trajectory Optimization Problem (UAVRTOP) that considers the kinematics and dynamics constraints of UAVs. Both surveys [10, 11], proposed taxonomy and classification of recent studies in terms of routing problems, their variants, approaches used, applications, etc. So that, the reader can have a general idea about the existing variants of routing problem and trajectory optimization for UAVs.

The reader may also refer to the survey in [12] that can provide a fast point of entry into the topic of UAVs operations planning for civil applications. In this survey, the authors summarized the important UAVs characteristics related to their operation planning, described planning problems for UAVs operations, and also planning problems for combined UAVs and other vehicles operations. They finally identified a number of new UAVs optimization problems and discussed model extensions from the well-known optimization problem such as TSP. However, all these surveys [10, 11, 12] are limited to only UAVs context.

In our paper, we propose a different survey that is complementary to the above cited surveys, since we dedicated this study to MTSP.

More precisely, our survey proposes the following contributions:

- Presentation of the most important real-life applications of MTSP.

- Analysis of the different MTSP existing variants and their formal description.

- A comprehensive review of recent contributions for ground and flying vehicles while highlighting the application area and the approach used to solve the MTSP for each contribution.

- A Taxonomy and a classification of recent contributions on MTSP that helps readers in their researches, and gives future research directions.

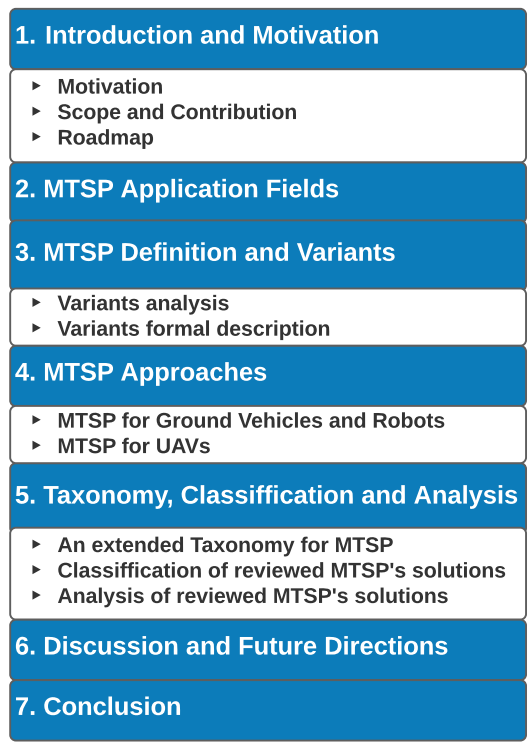

Figure 1: Paper Roadmap

The remainder of the survey is organized as follows. First, we mention in Section 2, the different real-life applications of MTSP. This motivates the reader about the reviewed optimization problem as we show that it is used to model several real-life applications. Then, we present in Section 3 an overview of MTSP, we analyze the main existing MTSP variants, and we give a formal description of them. After that, Section 4 is dedicated to review existing solutions proposed to solve MTSP for ground vehicles and robots in Section 4.1, as well for flying vehicles in Section 4.2. In Section 5, we propose a taxonomy, classification and analysis of the reviewed MTSP studies. Then, in Section 6 we discuss these reviewed contributions and give some future directions. Finally, we conclude the survey in Section 7. Figure 1 presents the roadmap of the present survey.

## 2. MTSP Application Fields

For years, mobile robots, vehicles, and UAVs, which are aircraft operating without a human pilot on board, have been considered as emerging technologies that have made many complex missions safer and easier. In order to achieve their missions, it is important to determine a path for each vehicle that optimizes a given objective while considering some constraints. MTSP has been adopted in different real-life applications to obtain optimized multiple vehicle routes. The main applications of MTSP are summarized in Figure 2 and are as follows:

***Transportation and Delivery***. such as good distribution or parcel delivery, or also bus transportation. Vehicles are in charge of transporting goods, parcels, or persons from a given location to another. In such an application, vehicle capacity and time constraints should be considered. In addition, a fleet of trucks can efficiently transport valuable goods, whereas drones can deliver small parcels in a very short time [13, 14, 15, 16, 17]. With the emerging of the UAVs technology, a new transportation service in which drones work in tandem with trucks to deliver parcels to customers is proposed to significantly reduce the delivery time.

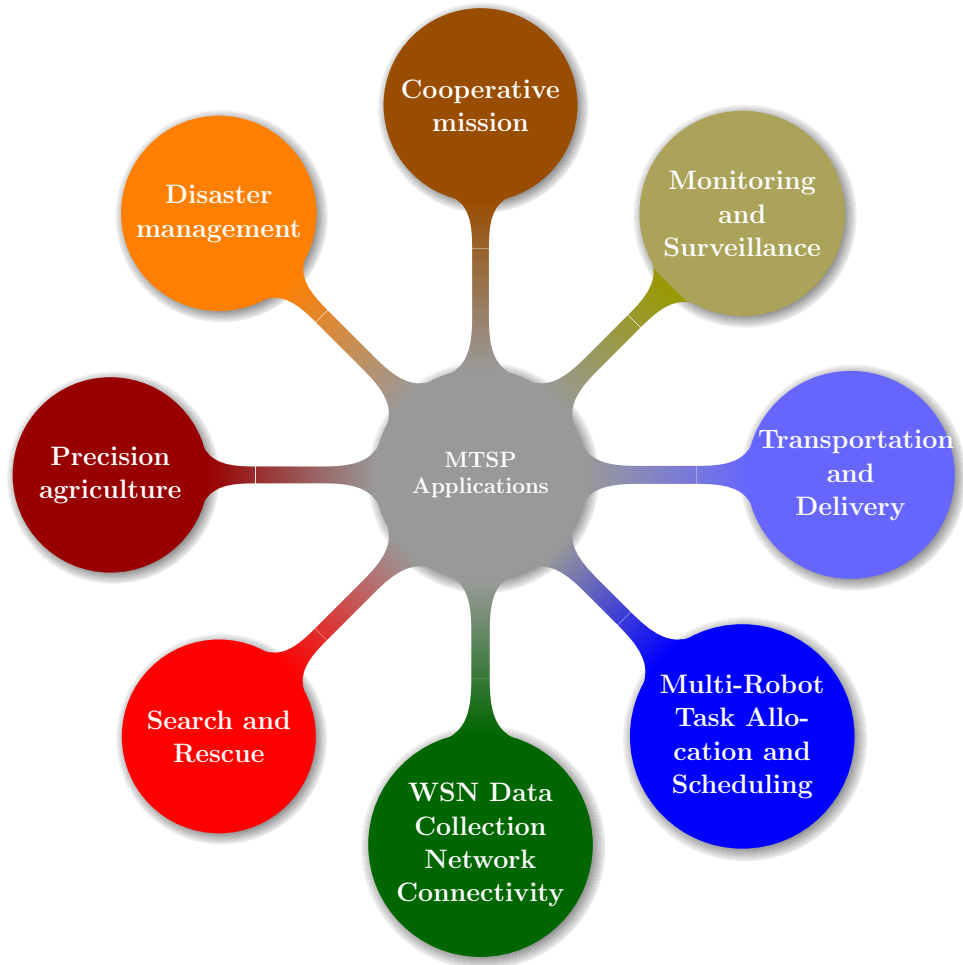

Figure 2: MTSP Applications

***WSN Data Collection and Network Connectivity***. in wireless sensor networks, additional sensor nodes may be deployed to forward the collected data until it reaches the sink. However, it is possible to use multiple mobile robots to act as a mobile sink and, therefore, help to collect data provided by sensor nodes. This strategy of data collection allows minimizing energy dissipation for sensor nodes located near the fixed sink and so increases the network lifetime [18, 19, 20, 21, 22]. The authors in [22], proposed a multi-objective optimization model for the joint problem of data collection and energy charging of sensor nodes. The proposed multi-objective model tries to optimize the total energy efficiency of the mobile robot and to reduce the average delay of data transmission of sensor nodes. In Delay Tolerant Network, mobile robots can re-establish network connectivity by picking up data from a source node and then delivering it to the destination node. In such an application, a fleet of UAVs can be used instead of ground robots. The idea is that these unmanned vehicles fly over disconnected ground nodes in charge of collecting data and act as a relay [23, 24, 18]. Constraints on storage capacity and data latency should be considered.

***Search and Rescue***. when human lives are at risk, it is essential to optimize every second during the search and rescue operation [25, 26]. In such an application, the locations to visit are determined then the routes to follow are optimized. With the emergence of the unmanned aerial vehicles technologies, UAVs are becoming very practical and useful for search and rescue operations [26].

***Precision agriculture***. mobile robots are widely used in precision agriculture, to ensure, for instance, crop monitoring or irrigation management. The farmer needs to provide an optimized path for each robot to reduce costs and improve production. Both ground vehicles and UAVs are of great help to farmers [27, 28, 29]. The authors in [29], proposed a cloud-based architecture composed of a WSN, mobile robots, and Cloud system to monitor a greenhouse region. The authors first generate candidate region monitoring points to be visited by mobile robots. Then, they generate the moving path of mobile robots to reach these points. To compute an optimal robots' path, the TSP and MTSP problems are used.

**Disaster management**. after a forest fire, an earthquake, or an industrial accident, mobile robots could help rescue teams in their mission [30, 31]. For example, the UAV's routes should be optimized in critical sites [32], such as during fire-fighting operations. The authors in [30] have proposed a cloud-based disaster management system. The system consists of a WSN deployed across an area of interest that will be monitored, and in case of a disaster, sensors nodes will report this information to a central station located at the cloud side. An optimized rescue plan is modeled as an MTSP and then generated using the Analytical Hierarchy Process MTSP (AHP-MTSP) method [31].

**Monitoring and Surveillance**. large areas could be monitored by mobile robots instead of sensor nodes. UAVs have been widely used in monitoring and surveillance applications. Most of the aerial vehicles used in such applications are of small UAVs platforms. However, those UAVs have limited energy. In such a case, either the UAV tour must not exceed its maximum energy or the UAV may make one or several refueling stops during its monitoring mission [33].

***Multi-Robot Task Allocation and Scheduling***. In the robotic community, several work have modeled the Multi-Robot Task Allocation (MRTA) problem as an MTSP. Generally speaking, the MRTA consists of allocating a set of tasks to each robot while optimizing some given metrics [9, 34, 35, 36, 37, 38].

***Cooperative mission***. when a swarm of vehicles/robots cooperates to accomplish a given mission such as target attacks in military applications [32]. In a cooperative mission, the computation of a vehicle's route should take into account the other vehicle' routes as well as collision avoidance. The trajectory of these vehicle must be optimized so that each vehicle can carry out its mission effectively, safely, and successfully. Notice that, in a cooperative mission, a cite may be visited several times by different robots. Some of the applications cited above may require cooperative robots such as disaster management [32] or parcel delivery using trucks and drones [13, 14, 15].

## 3. MTSP Definition and Variants

MTSP is widely studied and was originally defined as which, given a set of cities, one depot, $m$ salesmen and a cost function (e.g. time or distance), MTSP aims to determine a set of routes for $m$ salesmen minimizing the total cost of the $m$ routes, such that, each route starts and ends at the depot and each city is visited exactly once by one salesman.

MTSP has been applied to different application domains, which gave rise to new variants of this optimization problem. In this section, we analyze the different variants of MTSP, and then we provide a formal description of the main ones.

*3.1. Variants analysis*

In this paper, we extend MTSP variants presented in [9] and provide an analysis of more metrics based on recent contributions on MTSP. The new variants of MTSP result from considering the different characteristics of the salesmen, the depot, the city, and the problem constraints and objectives.

*Salesmen characteristics:.*

- The salesmen's type: depending on the application, the travelling salesmen in MTSP can be salesmen, vehicles, robots, or Unmanned Aerial Vehicles (UAVs) known also as drones.

  Note: In what follows, we use the term salesmen, vehicles and robots indifferently. Also, the term UAVs, drones, and flying vehicles are used indifferently.

- The number of salesmen $m$ is strictly greater than 1; otherwise, the problem is the same as TSP. $m$ can be fixed or to be determined by MTSP.

- Cooperative salesmen: several salesmen may cooperate to accomplish the given mission. They could be of the same vehicle type, or they may be combined, such as for delivery applications where trucks and drones work in tandem to deliver parcels to customers.

*Depot specifications:.*

- Single depot vs. Multiple depots: in the standard version of MTSP, only one depot should be considered, and its position is fixed. Since

several salesmen are used, the existence of multiple depots could optimize the cost of the tours. In such a variant of MTSP, the salesman can start its tour from a depot and join a different depot when it finishes its mission.

- Fixed vs. Mobile depot: generally, in MTSP the depot is fixed. However, in some applications, the depot can also be mobile. For example, a mobile depot can be a truck from which UAVs start and end their tours.

- Closed vs. Open path: in the classical MTSP, the salesman's path is closed since they have to start and end their tour from/at the same depot position. However, in some applications, the salesman does not need to return to depot and it can stay in the last visited city. Moreover, if multiple depots are considered, the salesman can join any of the existing depots that may be different from its initial depot.

- Refueling point: when the refueling is allowed, it can be performed either at the depots or at some additional refueling positions.

*Cities specifications:*.

- Standard MTSP: in the standard MTSP, all salesmen share the same workspace, i.e, they share all cities between them.

- Colored MTSP: a new variant introduced by [39] called colored traveling salesman problem (CTSP), where there are two groups of cities. One group shared between all salesmen and one group of cities that need to be visited by a specific salesman.

*Objective function:.*

- Single objective vs. Multi-objective : MTSP can be used to optimize a single objective or multiple objectives. Moreover, the main objectives addressed in the literature are:

  - Minimizing the total cost, in terms of (cumulative) distance or time, of all tours.
  - Minimizing the maximum salesman's tour cost.
  - Minimizing the mission time.
  - Minimizing the energy consumption.
  - Minimizing the number of salesmen.
  - Minimizing the additional cost such as the cost related to refueling stops.

It is worth noting that, the energy consumption and mission completion time are highly dependent on the distance traveled, and therefore, the main objectives considered in the literature are: the total distance traveled, the maximum robot's tour, and the balance between robots' tour length.

*Problem constraints:.*

- Energy constraint: the salesman consumes energy when moving. Vehicles like trucks do not have any constraint on their energy consumption, whereas other vehicles such as small UAVs have limited autonomy. Hence, if the vehicle cannot perform refueling during its mission, the computed tour will be constrained by the limited range of this vehicle.

- Capacity constraint: during its mission, the salesman may carry parcels or data. The same as for the energy constraint, the vehicle capacity constraint is generally considered with small vehicles such as UAVs, which can only carry small parcel and may have limited data memory storage.

- Time window constraint: it corresponds to the time interval during which the salesman needs to visit a given target. In some applications, it may correspond to data latency, where, the salesman has to pick up data from one site, carried it, and then, deliver it to another site.

*3.2. Variants formal description*

As previously explained, MTSP has several variants. In this section, we give a formal description of the main ones. For this purpose, we consider a set of $n$ Points of Interests (PoIs) or targets, saying $\{T_1, ..., T_n\}$ and a set of $m$ robots $\{R_1, ..., R_m\}$. The robots mission is to cooperatively visit these targets with an optimum cost.

According to the objective function, two main formulations of MTSP generally exist, which are [31]:

1. **MinSum MTSP**: in this MTSP variant, the objective function consists in minimizing the sum of all robots' tour costs. Formally speaking, the MinSum variant is modeled as:

$$\underset{Tour_{R_i} \in TOURS}{minimize} (\sum_{i=1}^{m} C(Tour_{R_i}))$$
$$\text{subject to}: Tour_{R_i} \cap Tour_{R_j} = \emptyset, \forall i \neq j, 1 \leq i, j \leq m. \quad (1)$$
$$\underset{i=1}{\overset{i=m}{\cup}} Tour_{R_i} = \{T_j, 1 \leq j \leq n\}.$$

where $C(Tour_{R_i})$ is the tour cost of robot $R_i$ and $TOURS$ is the set of all possible Tours. Moreover, the two conditions in Equation 1 guarantee that all targets are visited, and that each target is visited by only one robot.

2. **MinMax MTSP**: in this variant, the objective function consists in minimizing the longest tour cost (such as in terms of distance or time) among all robots' tours. This objective is highly adopted by studies focusing, for example, on the mission completion time. Formally speaking, this variant is modeled as:

$$\underset{Tour_{R_i} \in TOURS}{minimize} (\underset{j \in 1...m}{\max} \{C(Tour_{R_j})\})$$

$$\text{subject to}: Tour_{R_i} \cap Tour_{R_j} = \emptyset, \forall i \neq j, 1 \leq i, j \leq m \quad (2)$$

$$\underset{i=1}{\overset{i=m}{\cup}} Tour_{R_i} = \{T_j, 1 \leq j \leq n\}.$$

The $MinSum$ is mostly used when the total travelled distance or the robots' energy consumption should be minimized, whereas, the $MinMax$ is mostly used when the mission completion time must be minimized. Other objective functions could be derived as linear combinations of the ones above [40].

Moreover, the robots tour cost $C(Tour_{R_i})$ formulation varies whether the robots start from a single depot or they are initially located at different depots, and whether the robots need to return to their initial depot (closed path variant) or they can stop at the last visited target (open path variant).

More precisely, let $C(T_i, T_j)$ be the cost of traveling from target $T_i$ to target $T_j$, and $C(Tour_{R_i})$ be the tour cost of robot $R_i$, the robots tour cost in each variant can be modeled as follows:

**Single depot, closed path MTSP** : In this variant, all robots start from the same depot, and they return to it once finishing their mission. Therefore, the robot $R_i$ starts its tour from the depot $D$, then visits the list of $r$ assigned targets $\{T_{i_1}, ..., T_{i_r}\}$ in that order, and finally return to $D$. The tour cost of robot $R_i$ is equal to:

$$C(Tour_{R_i}) = C(D, T_{i_1}) + \sum_{k=1}^{r-1} C(T_{i_k}, T_{i_{k+1}}) + C(T_{i_r}, D) \qquad (3)$$

**Single depot, open path MTSP** : In this variant, the robot stop at the last visited target. The tour cost of robot $R_i$ is equal to:

$$C(Tour_{R_i}) = C(D, T_{i_1}) + \sum_{k=1}^{r-1} C(T_{i_k}, T_{i_{k+1}}) \qquad (4)$$

**Multiple depots, closed path MTSP** :

In this variant, the $m$ robots are initially located at $m$ different depots, saying $\{D_1, ..., D_m\}$. Therefore, the tour cost of robot $R_i$ is equal to:

$$C(Tour_{R_i}) = C(D_i, T_{i_1}) + \sum_{k=1}^{r-1} C(T_{i_k}, T_{i_{k+1}}) + C(T_{i_r}, D_i) \qquad (5)$$

**Multiple depots, open path MTSP** : In the open path variant of the multiple depots MTSP, the robots do not have to return to their initial depots and therefore, the tour cost of robot $R_i$ is equal to:

$$C(Tour_{R_i}) = C(D_i, T_{i_1}) + \sum_{k=1}^{r-1} C(T_{i_k}, T_{i_{k+1}}) \qquad (6)$$

For more details about MTSP formal description and MTSP possible formulations we refer the reader to [31] and [9].

In what follows, we identify two broad classes of MTSP, namely: (1) MTSP for salesman, vehicles, and robots, and (2) MTSP for UAVs. Indeed, UAVs based optimization problems have different characteristics and constraints than classical vehicles and robots.

## 4. MTSP Approaches

As its name suggests, MTSP was originally designed to optimize traveling salesman problems. It was then generalized to handle optimization tasks of ground vehicles or robots. However, over recent decades, UAVs, the new flying vehicle, have emerged. UAVs were first used in dangerous military missions to ensure the safety of pilots. Subsequently, these flying vehicles have attracted a great deal of interest in various civilian applications, and their use continues to grow. In the literature, several recent studies have proposed different approaches to solve MTSP. Most of the approaches proposed for flying vehicles are extended approaches of those applied for ground vehicles while considering UAVs constraints such as energy consumption or their limited carrying capacity.

In this section, we propose to study and analyse the different approaches applied to solve MTSP for ground vehicles (including salesmen and robots) and flying vehicles separately, in order to help the reader to understand the issues of each type of vehicles and to guide him/her in choosing the most appropriate approach to his/her optimization problem. These approaches are shown in Figure 3.

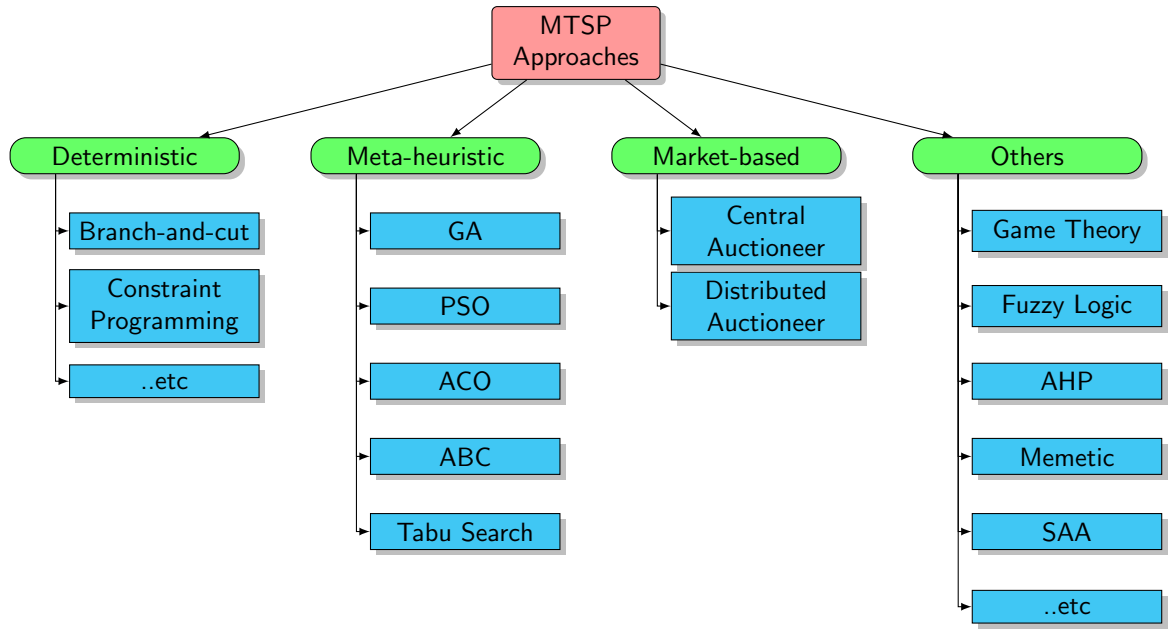

Figure 3: MTSP Approaches Taxonomy

*4.1. MTSP for Ground Vehicles and Robots*

This section is dedicated to review existing contributions that focused on MTSP for ground vehicles, robots, or salesmen in the general context. We have classified these solutions according to the used optimization approaches, as shown in Figure 3, and they are as following:

*4.1.1. Deterministic approaches*

Also known as exact methods which are able to reach to optimal solution for a given optimization problem. In general, these methods are very time-consuming due to the complexity of the calculations. Therefore, exact approaches are adopted in only few papers. The authors in [41], transformed the Multiple Depot Multiple Traveling Salesman Problem (MDMTSP) into a Single Depot Asymmetric TSP problem. In this paper, the cost of traveling between targets must satisfy the triangle inequality. To transform the problem from a multiple depot to a single depot MTSP, a set of extra nodes is added such that each new node is considered as a depot. Once the transformation is done, standard TSP exact methods are applied to solve the problem.

MDMTSP with heterogeneous vehicles has also been solved in [42] using an exact algorithm. The authors first introduced the integer linear programming (ILP) formulation. Then, they proposed a customized branch-and-cut algorithm that reached the optimal solution within 300 seconds using an instance of 100 targets and 5 vehicles.

Another study in [43] used constraint programming (CP) to formulate and to optimally solve MTSP by applying global constraints, interval variables, and domain filtering algorithms. However, the proposed approach is time-consuming since the execution time was more than 2 hours to solve an instance of 51 cities and 3 salesmen.

*4.1.2. Meta-heuristic based approaches*

The most used meta-heuristic algorithms in the literature are; Genetic Algorithm (GA), Ant Colony Optimization (ACO), Practical Swarm Optimization (PSO), and Artificial Bee Colony (ABC) techniques.

**GA based approaches**. The principle of GA techniques is based on the natural selection and genetics to generate the optimum solution from previous generation. The process starts from an initial random solutions (generations). Then, a fitness function is computed to evaluate the performance of the so-lutions at each iteration. After that, the process selects two parent solutions and computes a crossover/mutation operation to produce two new solutions which will be inserted in the next generation. If the computed child solutions have better fitness values, they will replace the parent solutions. This process of selection, crossover, and mutation is repeated until the new generation reaches the population size. This completes one iteration (generation). The algorithm continues until the number of certain generations (decided by user) is reached or the solution quality is no longer improved

In the GA approach a solution is represented as a chromosome. There are different chromosome representation techniques. The main ones used in the literature are summarized in Figure 4.

The authors in [45] proposed a two-part chromosome coding (Figure 4d) and developed a new crossover method. The proposed crossover method was compared to existing ones, such as the ordered crossover operator (ORX), the cycle crossover operator (CX), and the Partial matched crossover operator (PMX). The considered objectives are optimizing both the total distance traveled and the maximum tour length. The performance evaluation proved the efficiency of the proposed crossover method in generating a better solution quality.

The authors in [46] aims to build a genetic algorithm solution to solve MTSP. They first compared six different crossover operators, namely the cycle crossover (CX), partial matched crossover, order crossover (OX), edge recombination crossover (ERX), alternating position crossover (AEX), sequential constructive crossover (SCX). Experimental analysis on TSPLIB benchmark instances [47] of various sizes showed the performance of the

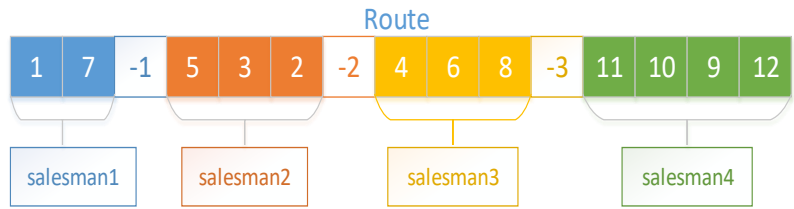

(a) Example of the one chromosome encoding

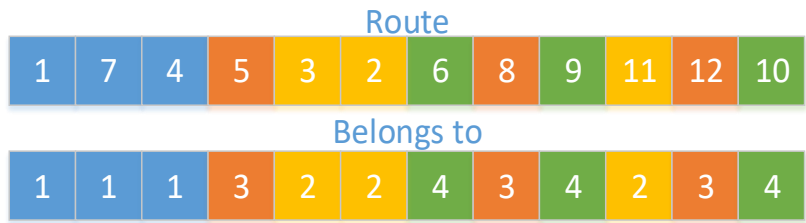

(b) Example of the two chromosomes encoding

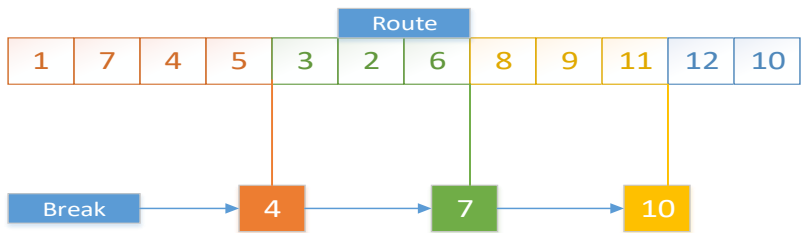

(c) Example of the two-part chromosome encoding (with breaks)

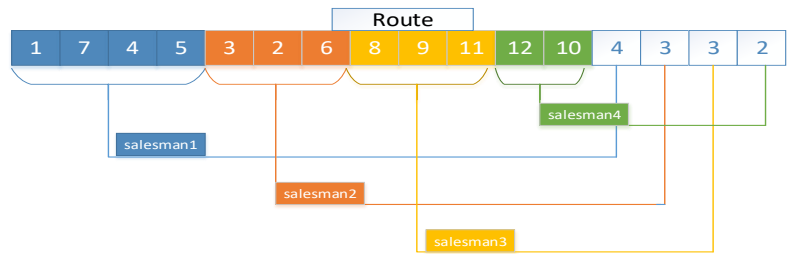

(d) Example of the two-part chromosome encoding (with cities per salesman)

Figure 4: Example of chromosome representation (12 Cities and 4 salesmen) in GA solutions [44]

crossover methods. Moreover, the experimental study showed that the sequential constructive crossover outperforms other crossover methods.

Recently, some studies focused on Partheno Genetic Algorithms (PGA) [48, 44]. The study in [48] introduced two partheno genetic algorithms. The first one is PGA with roulette and elitist selection, and it proposed four new types of mutation operations. The second one is called IPGA, and it proposed to bind together selection and mutation, where a wider mutation operator are used. The authors used a sequence encoding method to describe the real population by considering robot's route and breakpoint in the chromosome (Figure 4c). This chromosome is divided into parts: the first part represents the robot path and the second part corresponds to the breakpoint portion. The solutions were compared with a PSO solution for based on specific TSPLIB benchmarks [47]. Simulation results showed that IPGA has the best performance.

The authors in [44], studied the advantages and drawbacks of PGA in solving MDMTSP and reported the defect resulting from the lack of local information of individuals in the population. To resolve this defect, the authors proposed to integrate the reproduction mechanism in the Invasive Weed Algorithm (IWO). The resulting algorithm, called RIPGA for the Improved Partheno-Genetic algorithm with Reproduction mechanism, was compared with GA solutions to prove its efficiency in avoiding local convergence.

Some real applications need to optimize more than one criteria such as minimizing the robot's energy consumption, the mission completion time, etc. Therefore, the approach used to solve MTSP may need to optimize several objectives simultaneously. This kind of problem is known as the Multi-objective MTSP (MOMTSP). Indeed, the authors in [49] used the Non-dominated Sorting Genetic Algorithm (NSGA-II) to provide a solution to the multi-objective MTSP. The solution aims to optimize two objectives, namely: minimizing the total distance traveled and balancing the traveling times between the salesmen. The solution computes a set of non-dominated solutions. However, the authors did not explain how to compute the traveling time.

One more study in [50] proposed a multi-objective optimization problem called Multi-Robot Deploying wireless Sensor nodes problem (MRDS), where multiple robots have to deploy sensor nodes in given positions. In MRDS three objectives are considered; minimizing the mission time (i.e. the sensors deployment time), minimizing the number of used robots, and balancing robots tours time. The authors solved MRDS problem based on the NSGA-II algorithm.

***PSO based approaches***. The particle swarm optimization approach is one of the best-known meta-heuristics and has many similarities with genetic algorithms. PSO first initiates a population of random solutions, then it updates the generation until an optimum is reached. However, unlike the genetic algorithm, PSO does not use crossover or mutation operations to evolve the population. In PSO, possible solutions, known as particles and characterized by their velocity, move through the problem space following the current optimal particles.

The particle swarm optimization approach was applied in the study proposed in [51]. In this study, the authors addressed the problem of cooperative multi-robot task assignment and formulated it as MTSP. The solution aims to minimize both the total distance traveled and the maximum robot's tour cost. The authors extended the standard PSO to deal with several objectives. For that purpose, the author proposed two strategies, namely a Pareto front refinement strategy, which deletes inferior solutions, and a probability-based leader selection strategy. The authors compared the proposed approach with well known existing multi-objective approaches such as OMOPSO[52], SPEA2[53], NSGA-II[54], and SMPSO[55] and proved its superiority.

In the same context, the authors in [56] also addressed the MRTA problem and proposed DDPSO for dynamic and distributed PSO. The solution consists of two phases. In the first one, it groups tasks into clusters. Then, in the second phase, it assigns clusters to robots. The proposed solution was compared to distributed PSO and GA.

**ACO based approaches**. The Ant colony optimization approach is a population-based meta-heuristic to solving combinatorial optimization problems. ACO is inspired by the capability of real ant colonies to efficiently organize the food-seeking behavior of the colony by using chemical pheromone-tracks to communicate between ants. In the population, each individual is an artificial agent that progressively and stochastically constructs a solution to a given optimization problem. In the following, we review contributions based on the ACO approach to solve MTSP.

The authors in [57] have modeled the task assignment problem for multiple unmanned underwater vehicles as a constrained MTSP. The solution aims to minimize the total traveled distance and the total turning angle, which result in minimizing the vehicles energy consumption. Moreover, the solution enforces that the number of targets per vehicle must not exceed one. The proposed solution consists in two steps. First, the solution determines the number of targets to be assigned to each vehicle. Then, a proposed Multiple Ant Colonies System (MACS) method was used to solve the multi-objective MTSP. Experimental results proved that the proposed Multiple ACS outperforms the classical ACS.

In [58], the authors addressed the bi-criteria MTSP and adapted the ACO to solve the multi-objective single depot MTSP.

In [59], the authors proposed the mission-oriented ant team ACO (MOAT-ACO) algorithm. The MOATACO algorithm aims to minimize the total distance and to achieve load balance. In this solution, the authors endowed ants with directional awareness and loyal-to-duty characteristics, and proposed a mission pheromone to mimic a biological instinct. As a consequence, this relieved the entanglement between ants tours, and thus minimizes one of the objective, namely the total distance. To reach the second objective namely the load balance between tour, an ant firing rule was introduced, which allows the slowest ant to join harder working ones.

The authors in [60] addressed the MRTA problem and proposed an ACO based algorithm for multi-objective MTSP. The ACO method was integrated with a sequential variable neighborhood descent to further improve the Pareto set solutions locally. The solution aims to minimize both the total distance traveled and the maximum tour length simultaneously, and was compared to NSGA-II, MOEA/D-ACO [61] and FL-MTSP [62]. Although the solution outperforms these methods in terms of solution quality, it is slower than FL-MTSP. The authors in [63] addressed the MTSP with time window and proposed a hybrid solution by integrating ACO with the minimum spanning 1-tree to provide the optimal solution.

***ABC based approaches***. Like the ACO approach, the artificial bee colony is an optimization method focused on the honey bee swarm 's intelligent food-seeking behaviour. ABC is also a population-based method in which the location of a food source (i.e. population) is a potential solution to the optimization problem and the quantity of nectar from a food source represents the fitness of the associated solution.

Venkatesh et al. [64] used the ABC algorithm to solve the single depot MTSP aiming to minimize the total traveled distance (MinSum), and the maximum traveled distance (MinMax). The authors proposed also to use a local search to improve the obtained results. The same authors addressed the colored MTSP in [65] and also proposed an ABC based solution. A further study on the colored MTSP was addressed in [66]. The authors modified the ABC algorithm and introduced the generating neighbourhood solution (GNS) to solve MTSP.

***Hybrid approaches***. In the literature, several studies have proposed hybrid algorithms that combine different meta-heuristics and techniques to more effectively solve the multiple traveling salesman problem.

For instance, the study in [61] proposed a hybrid approach combining the ACO algorithm with the multi-objective evolutionary algorithm based on decomposition (MOEA/D) to resolve the multi-objective MTSP. The idea of the solution is to divide the multi-objective MTSP into several mono-objective sub-problems. Then, each mono-objective sub-problem is assigned to an ant. The ants are organized into groups, and each one has several neighboring ants. Each group is associated with a pheromone matrix, and each ant has a heuristic knowledge matrix. Moreover, each ant is charged with seeking the optimal solution to its assigned sub-problem. For that reason, the ant utilizes its heuristic knowledge matrix, its group pheromone matrix, and its current solution. The critical problems surrounding this method are the ambiguity of time convergence and the difficulty of implementation.

A new hybrid algorithm for large scale MTSP was proposed in [67]. The proposed solution, called AC-PGA for Ant Colony-PGA, is obtained by integrating PGA and ACO. More precisely, the algorithm first utilizes PGA to determine the best value of the salesmen's depots and the number of cities to be visited by each salesman. Then, it exploits ACO to compute the shortest path for each salesman.

Further study in [68] addressed the scheduling and routing of caregivers in a home health-care problem and formulated it as an MTSP with time window. They proposed a hybrid method combining ACO with memetic algorithm [69]. The solution aims to minimize the total traveling time and to balance the working time of caregivers (salesman).

*4.1.3. Market based approaches*

Market-based approaches, also known as auction-based, can be either centralized with a central auctioneer that receives bids and assigns the cities to the salesman with the lowest cost, or distributed where the bidding process is shared between the different salesmen.

The authors in [70] proposed a market based distributed and dynamic algorithm for MTSP, where the salesmen are robots. Thus the problem is called Multiple Traveling Robots Problem (MTRP). In this solution, each robot chooses its own targets in a progressive and distributed manner, as follows. Firstly, each robot uses the shortest distance as a cost function to select the appropriate targets. Then, it declares a single-item auction of its target visiting schedule. Choosing the best robot for a given task is done thanks to an auction based protocol called CNP (for Contract Net Protocol). The results of the Webots simulation showed that the proposed solution is efficient in terms of scalability, total path length, and communication overhead.

In [71], the multi-robot task assignment problem (MRTA) was formulated as MTSP, and solved using the K-means clustering technique with an auction process. The solution aims to optimize the total distance traveled and to balance the robots tours' length. First, $n$ groups of tasks are formed using the K-means algorithm, where $n$ is the number of robots. Then, each robot computes the cost of visiting each cluster formed in the previous step. Finally, in the auction step, the robots bid on clusters, and each cluster is assigned to the robot with the lowest cost. However, the algorithm's complexity is fairly high because it considers all possible combinations of cluster-robots assignments. Therefore this approach might not be appropriate to solve large scale instances.

Inspired by the Consensus Based Bundle Algorithm (CBBA) [72], and the Market Based Approach with Look-ahead Agents (MALA) [73], the authors of [74] introduced a market-based solution for MTSP. The solution is an repetitive market procedure, where robots perform the following steps in each iteration: market auction, agent-to-agent trading, agent switch, and agent relinquish step. Every robot selects the best tasks in the market auc-tion stage, based on the objective cost. Robots arbitrarily examine the tasks of other robots in the agent-to-agent trade step to verify if they can execute any of these tasks at a lower price. During the agent switch phase, the algorithm attempts to explore solutions that are not in the minimum of locality. The algorithm ends after a number of iterations without performance improvement.

Cheikhrouhou et al. proposed a market-based approach, called Move-and-Improve, in [75, 76]. The proposed solution involves the robots cooperating in allocating targets and incrementally eliminating possible overlap. The concept is simple: a robot moves and tries to optimize its solution at every step while communicating with its neighbours. Move-and-Improve approach consists of four main phases: (1) initial target allocation, (2) tour construction, (3) negotiation of conflicting targets, (4) solution improvement. Move-and-Improve was simulated using MATLAB and the Webots simulator and deployed on real robots using the Robot Operating System (ROS).

Another clustering market-based algorithm (CM-MTSP) [77] was proposed for the multi-objective MTSP. The solution consists of three steps, namely: clustering, auctioning, and improvement. In the clustering step, a central server group targets into clusters using k-means. Then, the server announces the already formed cluster one by one and robots bids on each cluster by computing the cost and sends it to the server. The server finally assigns the cluster to the robot with the lowest cost. The improvement step aims to optimize another objectives, namely the maximum distance traveled and the mission completion time. This improvement is achieved through the permutation of clusters between robots. In this algorithm, the authors assign a cluster to each robot. However, varying the number of clusters might lead to better results. Moreover, the authors tackled the multi-objective problem in a simultaneous way that might not necessarily give the best result.

*4.1.4. Other approaches*

In what follows, we review some MTSP contributions that adopted different techniques such as probability, game theory, fuzzy logic, analytical hierarchy process, etc.

To solve the MDMTSP, the authors introduced a method using probability collectives in [78], in which the vehicles are represented as autonomous agents and vehicles route as a strategy.

Khoufi et al. [79] proposed a multi-objective optimization problem to determine the robots tours responsible for collecting data from wireless sensor nodes and delivering this data to the depot. The proposed optimization problem should meet some constraints such as the data delivery latency, the robots energy, and the limited number of robots. This optimization problem is solved based on game theory approach. The proposed theory game is a coalition formation game that optimizes the maximum tour time, the number of robots, and balance the robots tours.

To address the multi-objective MTSP, a fuzzy logic-based solution (FL-MTSP) was introduced by Trigui et al. [62]. The solution considers two objectives, MinSum and MinMax. The FL-MTSP solution was compared with a GA based solution [34] to prove its efficiency.

Recently, Cheikhrouhou et al. [80, 31] proposed AHP-MTSP, an Analytical Hierarchy Process (AHP) based approach. AHP-MTSP first computes a weight for each considered objectives. These weights are computed based on user preferences and using the AHP method [81]. Then, the different objectives are aggregated into a single function, as a sum of the different weighted objectives. The comparison of AHP-MTSP to several methods including FL-MTSP [62] and CM-MTSP [77] proves its superiority.

Inspired by the work of [45], the authors in [82] suggested a Modified Two-Part Wolf Pack Search (MTWPS) method revised by the two-part chromosome encoding method and the transposition and extension operation for solving MTSP. The solution aims to minimize both the total distance traveled and the maximum tour.

Venkatesh et al. [64] proposed a meta-heuristic approach based on an invasive weed optimization algorithm. To further improve the solution, a local search method was also used.

Table 1 summarizes the different discussed MTSP solutions proposed for ground vehicles and robots.

Table 1: Summary of MTSP Solutions for Ground Vehicles and Robots.

| Class | Ref. | Objectives | Techniques | Description |
|---|---|---|---|---|
| Exact | [41] | MinSum | Transformation to TSP | The authors transformed a Multiple Depot, MTSP into a Single Depot Asymmetric TSP if the cost of the edges satisfy the triangle inequality. |
| | [42] | MinSum | branch-and-cut algorithm | The authors proposed an ILP formulation and then used a customized branch-and-cut algorithm |
| | [43] | MinSum | Constraint Programming | The authors formulated and solved MTSP using CP |
| Meta-heuristic | [45] | MinSum and MinMax | GA | The authors used a two-part chromosome encoding and introduced a new crossover operator |

Table 1: Summary of MTSP Solutions for Ground Vehicles and Robots.

| Class | Ref. | Objectives | Techniques | Description |
|---|---|---|---|---|
| | [46] | MinSum | GA | The study Considered six different crossover operators separately in order to find optimal solutions |
| | [48] | MinSum and MinMax | Partheno GA | Two partheno genetic algorithms (PGA). • PGA with roulette selection and elitist selection • IPGA, that binds together selection and mutation. |
| | [44] | MinSum | Partheno GA | The authors combine PGA with IWO and proposed RIPGA |
| | [49] | • MinSum • the salesmen's working times | NSGA-II | A multi-objective NSGA-II. |
| | [50] | • MinMax • the number of robots • the standard deviation between tours. | NSGA-II | A multi-objective MTSP called Multi-Robot Deploying wireless Sensor nodes problem (MRDS), where multiple robots have to deploy sensor nodes in a given positions. |
| | [51] | MinSum and MinMax | PSO | The algorithm extends the standard PSO to support multiple objectives. |
| | [38] | MinSum | PSO | The algorithm groups tasks into cluster and then assign cluster to robots |
| | [57] | • MinSum the total turning angle • While balancing the number of targets assigned to each vehicle | ACO | The algorithm first specifies the number of task then solve MTSP using ACO. |
| | [58] | • MinSum • Balancing Robots' tours | ACO | The algorithm used ACO to generate pareto front solutions. |

Table 1: Summary of MTSP Solutions for Ground Vehicles and Robots.

| Class | Ref. | Objectives | Techniques | Description |
|---|---|---|---|---|
| | [59] | MinSum | ACO | Four techniques were introduced in the search process of the ant teams: mission pheromone, path pheromone, greedy factor, and Max–Min ant firing scheme |
| | [60] | • MinSum<br>• MinMax | ACO | The ACO method was integrated with a sequential variable neighborhood descent to further improve the Pareto set solutions locally. |
| | [63] | MinSum | ACO | The ACO was improved by the minimum spanning 1-tree to solve MTSPTW. |
| | [64] | • MinSum<br>• MinMax | ABC | The authors solved SDMTSP by ABC and local search. |
| | [65] | MinSum | ABC | The authors addressed the CTSP. |
| | [66] | MinSum | ABC | The authors solved the large scale CTSP with ABC and GNS. |
| | [61] | • MinSum<br>• MinMax | • MOEA/D<br>• ACO | The algorithm is an integration of ACO with MOEA/D |
| | [67] | • MinSum<br>• With the constraint of minimum and maximum number of cities to be visited | • PGA<br>• ACO | First, PGA is used to determine the optimal salesmen's depots and number of cities per salesman. Then, ACO is used to compute the shortest route for each salesman. |
| | [68] | • MinSum<br>• Balancing the travelling Time | • ACO<br>• Memetic | The authors addressed the multi-objective MTSPTW using hybrid approach. |
| Market-based | [70] | MinSum | Contract Net Protocol (CNP) | An auction-based distributed and dynamic algorithm |
| | [71] | • MinSum<br>• Balancing the workload between the robots | • K-means for clustering<br>• Auction-based for allocation of clusters | To solve the balanced multi-robot task allocation problem it was modeled as an MDMTSP |

Table 1: Summary of MTSP Solutions for Ground Vehicles and Robots.

| Class | Ref. | Objectives | Techniques | Description |
|---|---|---|---|---|
| | [74] | • MinSum<br>• MinMax | Consensus-Based Bundle Algorithm | A market-based algorithm with iterative process |
| | [75, 76] | • MinSum<br>• MinMax | Move and improve | The algorithm involves the cooperation of the robots to allocate targets and remove possible overlap incrementally. |
| | [77] | • MinSum<br>• MinMax<br>• Minimizing the mission completion time | Clustering | The algorithm divides the locations into clusters and then allocates each cluster to the best robot. |
| Others | [78] | MinSum | Probability collectives | The authors used the Rosenbrock function in which the coupled variables are considered as autonomous agents working collectively to achieve the function optimum |
| | [79] | • MinMax Minimizing the number of robots<br>• Maximizing the fairness in term of tour duration | Game theory | A multi objective optimization problem to determine the robots tours in charge of collecting data from wireless sensor nodes and delivering this data to the depot. |
| | [62] | • MinSum<br>• MinMax | Fuzzy logic | The algorithm combines the objectives into a single fuzzy metric, transforming the problem to a single objective optimization problem. |
| | [80, 31] | • MinSum<br>• MinMax<br>• Balancing the tour length | AHP | The algorithm used the AHP technique to combine the objectives into a global function, and then optimize this global function. |

Table 1: Summary of MTSP Solutions for Ground Vehicles and Robots.

| Class | Ref. | Objectives | Techniques | Description |
|---|---|---|---|---|
| | [82] | ● MinSum<br>● MinMax | Two-part wolf pack search | The two-part wolf pack search algorithm was modified by the two-part chromosome encoding approach and the transposition and extension operation |

### 4.2. MTSP for UAVs

In this section, we review recent studies that used MTSP in a UAV context and provide a classification of these MTSP solutions according to the adopted approach while highlighting their application areas.

Table 2 summarizes the different reviewed MTSP solutions proposed for UAVs.

#### 4.2.1. Deterministic approaches

In transport and delivery application, the authors in [83] are the first to formally define the problem of combining an aerial vehicle with a truck for parcel delivery. Then, they extended their study in [13] to introduce the multiple flying sidekicks traveling salesman problem (mFSTSP), in which a truck and a fleet of UAVs are deployed to deliver small parcels to customers. This problem is formulated as a mixed-integer linear program (MILP) and solved via Gurobi for small size problems.

In the same context of parcel delivery, the authors in [14] were based on the study in [83] and proposed the Multiple Traveling Salesman Problem with Drones (MTSPD), which is based on both UAVs and trucks in the last-mile delivery. The proposed model is a variant of MTSP. In this optimization problem, multiple drones and multiple trucks perform deliveries together.

The objective of MTSPD is to minimize the arrival time of both vehicles, trucks and drones, at the depot after delivering all parcels to customers. To solve MTSPD, the authors proposed a mixed integer programming (MIP) formulation and obtained solution using IBM-CPLEX [84].

Due to computation complexity, the optimization problems introduced in [13] and [14] were solved optimally for only small instances. However, both studies applied heuristics to solve their problem for medium and large instances. We reviewed these heuristics in the next section.

Another study in [15] extended the problem of multiple traveling salesman problem proposed in [83]. In this study, a drone is in charge of dropping the parcel when it arrives at the customer and then it can either flies to another customer to pickup a new parcel or return directly to the depot to start new delivery. This problem is modeled as an unrelated parallel machine scheduling (PMS), and it integrates multiple depots with multiple trucks and drones, constrained by time-window, drop-pickup synchronization, and multi-visit. To solve this problem, a novel application of constraint programming approach is proposed in order to minimize the maximum time needed to satisfy all delivery tasks.

In the context of attacking multiple targets, the problem of cooperative trajectory planning integrated target assignment for multiple Unmanned Combat Aerial Vehicles (UCAV) is studied in [32]. The problem is formulated as dynamics-constrained, multiple depots, multiple traveling salesman problem with neighborhoods (DC_MDMTSPN), which is a variant of MTSP. The proposed problem take into account the battlefield environment constraints (e.g. threat avoidance) and the UCAV dynamics model (e.g. avoid collision with each other) and was solved based on a two-phases approach. In the first phase, the authors construct the directed graph and then transform the original problem into asymmetric TSP (ATSP). In the second phase, they use the Lin−Kernighan Heuristic (LKH) searching algorithm to solve the ATSP.

The study in [85] introduced a problem formulation of the heterogeneous multi-UAV task assignment problem as multiple traveling salesman problem. The proposed problem considers situations that include the traditional offline centralized situation and the offline with parameter uncertainty situation. The authors proposed a novel digraph-based deadlock-free algorithm as well as a Modified Two-part Wolf Pack Search algorithm to solve the deterministic offline problem efficiently.

*4.2.2. Meta-heuristic based approaches*

Meta-heuristics have been applied to solve many NP-hard optimization problems. In this section, we provide a review of contributions that solved MTSP based on a meta-heuristic such as Genetic Algorithm and Tabu Search. Notice that many studies proposed combined heuristics where the problem is solved following several phases and based on different techniques (e.g. clustering, meta-heuristic) to reduce the problem complexity.

**GA based approaches**. In the context of trucks and drones in the last-mile delivery, the authors in [14], solved optimally MTSPD for small size instances and also proposed a new heuristic called Adaptive Insertion algorithm (ADI) to solve large size instances. The principle of ADI is to first build an initial solution and second to improve it by applying removal and insertion operators to construct the MTSPD solution. To perform the second phase of ADI, the authors were based on the genetic algorithm, the combined K−means/Nearest Neighbor and the random Cluster/Tour. Simulation results showed that, in small size instances, both solvers and the genetic meta-heuristic, GA−ADI, reached the optimal solution. Notice that GA-ADI solved the MTSPD problem significantly faster than the solver IBM-CPLEX [84]. However, in large size instances, the authors used only the GA−ADI heuristic and they demonstrated the efficiency of using multiple UAVs to reduce the delivery time.

The authors in [23] proposed a study addressing UAVs as DTN relays, and introduced a proactive scheme called Deadline Triggered Pigeon with Travelling Salesman Problem with Deadlines (DTP-TSP-D). In this problem, UAVs can communicate with one or a cluster of nodes, pick up data from a ground node while considering it capacity of carrying messages from one location to another and then hovering until being triggered to deliver messages directed to other ground node. A genetic algorithm is used to determine the optimized UAVs tours in terms of delivery time.

A similar study in [24] proposed to use a UAV as a message ferry node that is in charge of traveling among disconnected nodes in a DTN network to deliver their messages. The authors introduced a multiple message ferry UAVs optimization problem which is a variant of MTSP so that the optimal path planning minimizes the message delivery delay. To obtain the optimal path planning solution, the genetic algorithm is adopted to solve the problem in a feasible time. In the genetic algorithm, clusters of nodes are built, each cluster will be assigned to a UAV, then the UAV's path visiting all nodes inside a cluster is determined based on the traffic flows between nodes and a load of messages nodes in those nodes in order to optimize message delivery delay in the network. In a small network, this genetic algorithm provides good solution as those obtained with an exhaustive search approach but in shorter run time. In addition, the proposed optimization problem outperforms the traditional MTSP solution in terms of message delivery delay in DTN.

The study in [18], addressed the problem of data collection using multiple mobile sinks in large scale WSN, in order to save energy throughout the network and to increase the packet delivery rate. To solve this problem, the authors proposed a two-phase heuristic. In the first phase, $k$ clusters of sensor nodes are formed, where $k$ is the number of UAVs acting as mobile sinks. In the second phase, the path of each UAV is determined based on a smooth path construction (SPC) algorithm. In SPC, the genetic algorithm is used to determine each UAV's tour visiting each sensor node inside the cluster exactly once. Then, the smooth tours are computed based on UAV's turning constraints.

When cooperative UAVs operate in hazardous areas to perform multiple tasks, optimizing the number of UAVs deployed, as well as the trajectory of each of them, would help to accomplish the mission in a minimum amount of time. In this context, the study in [86] proposed the Multi−UAVs to multi-tasks, task assignment and path planning problem which is a variant of MTSP that optimizes the number of UAVs for the given time constraint and task set. To solve this problem, a coordinated optimization algorithm combining the genetic algorithm and cluster algorithm, is introduced in this paper. The authors proposed first the K-means clustering algorithm to build clusters of multiple tasks, where each cluster will be assigned to one UAV. Second, adjacent tasks of the same cluster are grouped in order to reduce the number of locations to be visited by the UAV. After that, a TSP optimization problem will be solved based on the genetic algorithm while considering time constraints. The comparison evaluation between the coordinated optimization algorithm and GA showed that the proposed coordinated optimization algorithm is more effective than GA.

In the context of monitoring applications where UAVs are in charge of covering multiple regions, the study in [87] proposed the Energy Constrained Multiple Traveling Salesman Problem for Coverage Path Planning (EMTSP-CPP). To solve this problem, the authors introduced a modified version of the genetic algorithm while considering the new representation of the individual (chromosome), as well as a new version of crossover and mutation operation and, finally, a constraint-aware fitness function. The authors demonstrated that the modified genetic algorithm has a better performance compared to other approaches.

In [26], a multi-objective UAV path planning for search and rescue was proposed. The proposed solution is based on genetic algorithm and aims to minimize the completion time. The solution divides the environment into cells and, then, uses MTSP to guarantee that each cell is searched by exactly one UAV.

In precision agriculture, UAVs can be deployed to spray fields with pesticides. In this context, the study in [88] introduced the mission assignment and the path-planning problem with Multi-Quadcopters which are a specific type of UAVs. The proposed problem was formulated as the MTSP optimization with the objective is to reduce the mission completion time while considering a constraint on the Quadcopter's battery capacity limitation. To solve this problem, the authors proposed a hierarchal approach in which an inner-and-outer loop structure is employed. Indeed, the inner loop is based on the genetic algorithm while the outer loop uses a nonlinear programming method is based on the optimal results obtained by the inner loop. The efficiency of the proposed approach was illustrated based on performance comparisons to a conventional approach.

***TS based approaches***. The tabu search approach is a meta-heuristic based on the local search methods and used for mathematical optimization. Local search methods tend to be stuck in sub-optimal solutions. Therefore, Tabu search improves the performance of these methods by exploring the solution space beyond local optimality.

In the field of monitoring and surveillance application, the study in [33] proposed the two-stage Fuel-Constrained Multiple-UAV Routing Problem (FCMURP). The FCMURP is a generalization of the multiple traveling salesman problem and a variant of the vehicle routing problem for UAVs with fuel constraints. In FCMURP, multiple depots are considered, and they also represent refueling points. However, there is only one depot from where each UAV tour starts and ends. The two-stage FCMURP works as follows: In the first stage, MTSP is adopted to build a tour for each UAV. However, if any UAV does not have sufficient energy to complete its tour, it can make a refueling stop at any depot. In the second stage, additional refueling stops are added to the first-stage tours to ensure feasibility for the achievement of the energy consumption. In the first stage of FCMURP, the objective is to minimize the sum of the distance traveled by all UAVs whereas the objective of the second stage is to minimize the expected traveling distance for additional refueling stops. To solve FCMURP, the authors proposed the Sample Average Approximation (SAA) approach. In small instances, the SAA approach converges to optimal solutions for the two-stage model, however, in medium and large instances, the SAA takes so much time to converge. To cope with this problem, the authors proposed to solve FCMURP based on a tabu search-based heuristic and they showed that this heuristic provides high quality solutions.

*4.2.3. Other approaches*

In trucks and drones delivery application, the authors in [13] proposed a three-phase heuristic to solve the multiple flying sidekicks traveling salesman problem (mFSTSP), with 100 customers and 4 drones. In the first phase, some customers are selected to be served by the truck and others will be served by UAVs. Then, the truck tour is determined based on the traveling salesman optimization problem. In the second phase, each customer is assigned to a specific UAV and both truck and UAV tours are identified. In the third phase, the authors solved a MILP to determine the exact timing of the scheduled operations for the truck and the UAVs. The flight endurance for the heterogeneous UAVs is modeled as a function of their battery size, payload, travel distance, and flight phases. Finally, a local search procedure is executed to improve the solution quality. The authors showed that for problems of realistic size the three-phase heuristic provides high-quality solutions with reasonable execution time.

In the context of task allocation and scheduling, the authors in [85] formulated the multi-UAV task assignment problem as MTSP. The presented problem considers various situations, including multiple consecutive tasks, heterogeneous UAVs, time-sensitive, and uncertainty. To solve this problem with parameter uncertainty, several methods were introduced, such as robust optimization method, the duality theory and a novel combined algorithm, including the classical interior point method and the modified two-part wolf pack search algorithm. The authors also proposed an online hierarchical planning algorithm to solve the online problem with the time-sensitive uncertainty. Finally, they conducted several numerical simulations and physical experiments to check the efficiency of their presented algorithms.

Table 2 summarizes existing solutions to MTSP for UAVs.

Table 2: Summary of MTSP Solutions for UAVs

| Class | Ref. | Objectives | Techniques | Description |
| --- | --- | --- | --- | --- |
| Exact | [13] | MinMax (time) | Formulated as MILP and solved via Gurobi version 7.0.1 | A Truck and drones in the last-mile delivery. Due to the computation complexity the problem was solved optimally for small instances. |
| | [14] | MinMax (time) | Formulated as Mixed Integer Programming (MIP) and solver by Cplex solver | Trucks and drones parcel delivery for last mile. The problem was solved optimally for small instances. |
| | [15] | MinMax (time) | Modeled as PMS. Constraints programming approach | Trucks and drones in the last-mile delivery. The problem integrates:<br>• multiple depots,<br>• multiple trucks and drones,<br>• time-window's constraint,<br>• drop-pickup synchronization,<br>• multi-visit. |

Table 2: Summary of MTSP Solutions for UAVs

| Class | Ref. | Objectives | Techniques | Description |
|---|---|---|---|---|
| | [32] | MinMax (distance) | Transformation to ATSP | Cooperative mission for multiple target attack. The (DC_MDMTSPN) is solved based on two phases: <br> • the problem is transformed to ATSP, <br> • the Lin−Kernighan Heuristic (LKH) searching algorithm is used to solve the ATSP. <br> UAV's dynamic constraints are considered. |
| | [85] | Min (distance) | • Digraph-based deadlock-free algorithm <br> • MTWPS algorithm | Multi-UAV task allocation and scheduling. A novel digraph-based deadlock-free algorithm as well as Modified Two-part Wolf Pack Search MTWPS algorithm are proposed to solve the deterministic offline problem efficiently. |
| Meta-heuristic | [14] | MinMax (time) | ADI heuristic combined with: <br> • GA, <br> • K−means/ Nearest Neighbor, <br> • Random Cluster/Tour. | Trucks and drones in the last-mile delivery. The ADI's heuristic builds first an initial solution and second improves it by applying removal and insertion operators to construct the MTSPD solution. |
| | [24] | Minimizing the message delivery delay | GA | Message Delivery in DTN. <br> In the genetic algorithm: <br> • clusters of nodes are built, <br> • each cluster is assigned to a UAV, <br> • each UAV's tour is determined such that the message delivery delay is optimized. |

Table 2: Summary of MTSP Solutions for UAVs

| Class | Ref. | Objectives | Techniques | Description |
|---|---|---|---|---|
| | [23] | • Maximizing the delivery ratio • Minimizing the average delivery delay. | • GA • Proactive schemes: Message Ferrying and Homing Pigeon | Message Delivery in DTN. In the proposed scheme each UAV switches between two modes: • Ferry mode: each UAV follows a ferry route in its cluster of ground nodes. This route is computed based on GA. • Pigeon mode: each UAV delivers the Data collected following a route computed based on GA |
| | [18] | • Maximizing the delivery rate • Minimizing the energy consumption. | • Clustering • GA | Data collection using UAVs. Two-phase Heuristic • Clusters of nodes are formed • Construction of UAVs' smooth path inside each cluster based on Genetic algorithm. |
| | [86] | MinSum (distance) | • GA • Clustering | Cooperative UAVs operating in hazardous areas. The problems is solved as follows : • the K−means algorithm to build clusters of multiple tasks, each cluster is assigned to one UAV. • adjacent tasks of the same cluster are grouped to reduce the UAV path. • GA is used to determine the UAVs' tours. |
| | [26] | Minimizing the completion time | GA | A multi-objective UAV path planning for search and rescue missions is proposed and solved based on the genetic algorithm. |

Table 2: Summary of MTSP Solutions for UAVs

| Class | Ref. | Objectives | Techniques | Description |
|---|---|---|---|---|
| | [33] | • MinSum (distance) | • SAA<br>• Tabu Search | Monitoring and surveillance application. The two-stage FCMURP works as follows:<br>• first stage MTSP is adopted to build a tour for each UAV.<br>• second stage, additional refueling stops are added to satisfy energy consumption.<br>FCMURP is solved using:<br>• SAA for small instances<br>• Tabu Search for meduim and large instances |
| | [87] | • MinSum<br>• MinMax | GA | Covering multiple regions using UAVs. A problem called Energy Constrained MTSP for Coverage Path Planning (EMTSP-CPP) is proposed and solved using a modified Genetic algorithm. |
| | [88] | MinMax (time) | Hierarchal approach based on GA | In precision agriculture, Multi-Quadcopters are deployed to spray fields with pesticides. The limited battery capacity is considered.<br>• A hierarchal approach using inner-and-outer loop structure • The inner loop is based on the genetic algorithm<br>• The outer loop uses a nonlinear programming method |

Table 2: Summary of MTSP Solutions for UAVs

| Class | Ref. | Objectives | Techniques | Description |
|---|---|---|---|---|
| Others | [13] | MinMax (time) | Three-phase heuristic | Truck and drones in the last-mile delivery. The problem was solved for large instances based on 3 phases heuristic:<br>• phase 1: dividing customer's delivery tasks between the truck and the drones.<br>• phase 2: assigning each customer to a specific UAV and identifying truck and UAV tours.<br>• phase 3: determining the schedule of each truck and drone's activities. |
| | [85] | MinSum (distance) | • Hierarchical planning algorithm<br>• a novel combined algorithm including the classical interior point method and the MTWPS algorithm | Multi-UAV task allocation and scheduling. The problem to solve considers various situations including multiple consecutive tasks, heterogeneous UAVs, time-sensitive and uncertainty. |

## 5. Taxonomy, Classification and Analysis

In this section, we first provide an extended taxonomy for MTSP, which is based on the MTSP variants, the applied optimization approaches, and the application domains for which the solution was proposed. After that, the previously reviewed solutions are classified according to this proposed taxonomy. This classification presents an overview of the existing MTSP studies which can help the readers to select the suitable MTSP variant for a given application, as well as the approach used to solve the problem. Finally, we conclude this section by an analysis study of the reviewed papers, based on the metrics provided in our classification.

*5.1. An extended Taxonomy for MTSP*

In the following, we propose an extended taxonomy for MTSP. First, we start by enumerating the attributes considered in this taxonomy based on three criteria, namely: MTSP's variants, approaches, and application fields. We select the most common attributes for the MTSP variants which are previously described in Section 3. We also select the different approaches applied to solve MTSP (cited in Section 4). Moreover, the different MTSP application fields are previously detailed in Section 2.

- MTSP Variants

    - Salesman:
        * 1- Salesmen
        * 2- Robots
        * 3- Vehicles
        * 4- UAVs

    - Depot:
        * 5- Single
        * 6- Multiple
        * 7- Refueling point

    - Cities:
        * 8- Standard MTSP

- - - * 9- Colored TSP

  - – Problem Constraints:

    * 10- Energy

    * 11- Capacity

    * 12- Time window

  - – Problem Objectives:

    * 13- Single-objective optimization problem

    * 14- Multi-objective optimization problem

- Approaches

  - 15- Exact algorithm

  - 16- Genetic Algorithm (GA)

  - 17- Particle Swarm Optimization (PSO)

  - 18- Ant Colony Optimization (ACO)

  - 19- Artificial Bee Colony (ABC)

  - 20- Tabu Search (TS)

  - 21- Clustering algorithm (e.g. K−means, Nearest neighbor)

  - 22- Market-based

  - 23- Others

    * a- Fuzzy logic

    * b- Game theory

    * c- ADI heuristic

- * d- Three-phases heuristic
- * e- Sample Average Approximation (SAA)
- * f- Lin−Kernighan Heuristic (LKH)
- * g- Proactive scheme
- * h- Probability
- * i- Analytical Hierarchy Process
- * j- Two-Part Wolf Pack Search
- * k- Memetic

- Applications
  - 24- Transportation and Delivery
  - 25- Data Collection
  - 26- Search and Rescue
  - 27- Precision agriculture
  - 28- Disaster management
  - 29- Monitoring and Surveillance
  - 30- Multi-Robot Task Allocation and scheduling
  - 31- Cooperative mission
  - 32- General context

*5.2. Classification of reviewed MTSP's solutions*

We classify the existing studies reviewed in our survey according to the previously proposed taxonomy as illustrated in Table 3.

| Ref | Year | Variants | | | | | | | | | | | | | | Approaches | | | | | | | | | Applications | | | | | | | | |
|---|---|---|---|---|---|---|---|---|---|---|---|---|---|---|---|---|---|---|---|---|---|---|---|---|---|---|---|---|---|---|---|---|---|
| | | Salesman | | | | Depot | | | Cities | | Const. | | Obj. | | | | | | | | | | | | | | | | | | | | | |
| | | 1 | 2 | 3 | 4 | 5 | 6 | 7 | 8 | 9 | 10 | 11 | 12 | 13 | 14 | 15 | 16 | 17 | 18 | 19 | 20 | 21 | 22 | 23 | 24 | 25 | 26 | 27 | 28 | 29 | 30 | 31 | 32 |
| | | | | | | | | | | | | | | Ground Vehicles and Robots solutions | | | | | | | | | | | | | | | | | | | |
| [70] | 2007 | | ✓ | | | | ✓ | | ✓ | | | | ✓ | | | | | | | | | | ✓ | | | | | | | | | | ✓ |
| [57] | 2008 | | | ✓ | | | ✓ | | ✓ | | | | | ✓ | | | | | | | | | | | | | | | | | ✓ | | ✓ |
| [41] | 2009 | ✓ | | | | | | | ✓ | | | | | ✓ | ✓ | | | | | | | | | | | | | | | | | | ✓ |
| [78] | 2010 | | ✓ | | | | ✓ | | ✓ | | | | | ✓ | ✓ | | | | ✓ | | | | | h | | | | | | | ✓ | | ✓ |
| [71] | 2011 | | ✓ | | | | ✓ | | ✓ | | | | | | ✓ | | | | | | | | ✓ | | | | | | | | | | ✓ |
| [74] | 2011 | | ✓ | | | | ✓ | | ✓ | | | | | | ✓ | | | | | | | | ✓ | | | | | | | | | | ✓ |
| [45] | 2013 | ✓ | | | | | ✓ | | ✓ | | | | | | ✓ | | | | | | | | | | | | | | | | | | ✓ |
| [61] | 2013 | ✓ | | | | | ✓ | | ✓ | | | | | ✓ | ✓ | | ✓ | | | | | | | | | | | | | | | | ✓ |
| [75] | 2014 | | ✓ | | | | ✓ | | ✓ | | | | | | ✓ | | ✓ | | | | | | | | | | | | | | | | ✓ |
| [49] | 2015 | ✓ | | | | | ✓ | | ✓ | | | | | | ✓ | | ✓ | | | | | | | | | | | | | | | | ✓ |
| [50] | 2015 | ✓ | | | | ✓ | ✓ | | ✓ | | | | | ✓ | ✓ | | ✓ | | ✓ | | | | ✓ | | | | | | | | | | ✓ |
| [58] | 2015 | ✓ | | | | ✓ | ✓ | | ✓ | | | | | | ✓ | | | | | | | | | | | | | | | | | | ✓ |
| [64] | 2015 | ✓ | | | | | ✓ | | ✓ | | | | | | ✓ | ✓ | ✓ | | | | | | ✓ | | | | | | | | | | ✓ |
| [77] | 2016 | | ✓ | | | | ✓ | | ✓ | | | | | ✓ | ✓ | ✓ | | | | | | | | | | | | | | | | | ✓ |
| [79] | 2016 | | ✓ | | | | ✓ | | ✓ | | | | | ✓ | ✓ | | | | ✓ | | | | ✓ | b | | | ✓ | | | | | | ✓ |
| [62] | 2017 | ✓ | | | | | ✓ | | ✓ | | | | | ✓ | ✓ | | | | | ✓ | | | | a | | | | | | | ✓ | | ✓ |
| [82] | 2017 | ✓ | | | | | ✓ | | ✓ | | | | | | ✓ | | | | | | | | | j | | | | | | | | | ✓ |
| [42] | 2017 | | ✓ | ✓ | | | ✓ | | ✓ | | | | | | ✓ | | | | | | | | | | | | | | | | | | ✓ |
| [43] | 2017 | ✓ | | | | | ✓ | | ✓ | | | | | | ✓ | | | | | | | | | | | | | | | | ✓ | | ✓ |
| [31] | 2018 | ✓ | | | | | ✓ | | ✓ | | | | | | ✓ | | | | ✓ | | | | | i | | | | | | | | | ✓ |
| [48] | 2018 | ✓ | ✓ | | | ✓ | ✓ | | ✓ | | | | | | ✓ | | ✓ | ✓ | | | | | | | | | | | | ✓ | | | ✓ |
| [60] | 2018 | ✓ | | | | | ✓ | | ✓ | | | | | ✓ | | | ✓ | | | ✓ | | | | | | | | | | ✓ | | | ✓ |
| [65] | 2019 | ✓ | ✓ | | | ✓ | ✓ | | ✓ | ✓ | | | | | ✓ | | | | | | | | | | | | | | | | | | ✓ |
| [46] | 2019 | ✓ | | | | | ✓ | | ✓ | | | | | | ✓ | | ✓ | | ✓ | | | | | | | | | | | | | | ✓ |
| [38] | 2019 | ✓ | | | | | ✓ | | ✓ | ✓ | | | | | ✓ | | | ✓ | | | | | | | | | | | | | | | ✓ |
| [59] | 2019 | ✓ | ✓ | | | ✓ | ✓ | | ✓ | | | | | | | | ✓ | | ✓ | ✓ | | | | | | | | | | | ✓ | | ✓ |
| [66] | 2019 | ✓ | | | | | ✓ | | ✓ | | | | | | | | ✓ | | | | | | | | | ✓ | | | | | | | ✓ |
| [68] | 2019 | ✓ | | | | | ✓ | | ✓ | | | | | ✓ | | | ✓ | | ✓ | | | | | | | | | | | | | | ✓ |
| [44] | 2020 | ✓ | | | | | ✓ | | ✓ | | | ✓ | | ✓ | | | | | ✓ | | | | | k | | | | | | | | | ✓ |
| [51] | 2020 | ✓ | ✓ | | | | ✓ | | ✓ | | | | | ✓ | ✓ | | | | | | | | | | | | | | | | | | ✓ |
| [67] | 2020 | ✓ | | | | | ✓ | | ✓ | | | | | ✓ | ✓ | | | | | | | | | | | | | | | | | | ✓ |
| [63] | 2020 | ✓ | | | | ✓ | ✓ | | ✓ | | | ✓ | | ✓ | ✓ | | | | ✓ | | | | | | | | | | | | | | ✓ |

| Ref | Year | Variants | | | | | | | | | | | | | Approaches | | | | | | | | | | Applications | | | | | | | | |
| --- | --- | --- | --- | --- | --- | --- | --- | --- | --- | --- | --- | --- | --- | --- | --- | --- | --- | --- | --- | --- | --- | --- | --- | --- | --- | --- | --- | --- | --- | --- | --- | --- | --- |
| | | Salesman | | | | Depot | | | Cities | | Const. | | Obj. | | | | | | | | | | | | | | | | | | | | | |
| | | 1 | 2 | 3 | 4 | 5 | 6 | 7 | 8 | 9 | 10 | 11 | 12 | 13 | 14 | 15 | 16 | 17 | 18 | 19 | 20 | 21 | 22 | 23 | 24 | 25 | 26 | 27 | 28 | 29 | 30 | 31 | 32 |
| | | | | | | | | | | | | | | | | UAV solutions | | | | | | | | | | | | | | | | | |
| [18] | 2015 | | | | ✓ | ✓ | | | ✓ | | | | ✓ | | | ✓ | | | | | | ✓ | | | | ✓ | | | | | | | | |
| [32] | 2016 | | | | ✓ | | ✓ | ✓ | ✓ | | | | ✓ | | | ✓ | | | | | | | | f | | | | | | | | ✓ | | |
| [26] | 2017 | | | | ✓ | ✓ | ✓ | | ✓ | | | | ✓ | | | ✓ | ✓ | | | | | | | | | | ✓ | | | | | | | |
| [15] | 2018 | | | ✓ | ✓ | ✓ | | | ✓ | | | | ✓ | | | | ✓ | | | | | | | | ✓ | | | | | | | | | |
| [23] | 2018 | | | | ✓ | ✓ | | | ✓ | | | ✓ | ✓ | | | | ✓ | | | | | ✓ | | g | | | ✓ | | | | | | | |
| [24] | 2018 | | | | ✓ | ✓ | | | ✓ | | | ✓ | ✓ | | | | ✓ | | | | | ✓ | | | | | ✓ | | | | | | | |
| [33] | 2018 | | | | ✓ | ✓ | | ✓ | ✓ | | ✓ | | ✓ | | | | | | | | ✓ | | | e | | | | | | | | | | |
| [85] | 2018 | | | | ✓ | ✓ | | | ✓ | | ✓ | | ✓ | | | | ✓ | | | | | | | j | | | | | | | | | | |
| [88] | 2018 | | | | ✓ | ✓ | ✓ | | ✓ | | ✓ | | ✓ | | | | ✓ | | | | | | | | | | | | ✓ | | | | | |
| [86] | 2019 | | | | ✓ | ✓ | | | ✓ | | ✓ | | ✓ | | | | ✓ | | | | | | ✓ | | | | | | | | | | | |
| [87] | 2019 | | | | ✓ | ✓ | | | ✓ | | | ✓ | ✓ | | | ✓ | | | | | | | | j | | | | | | | | | | |
| [14] | 2019 | | | ✓ | ✓ | | | | ✓ | | | ✓ | ✓ | | | ✓ | | | | | | | ✓ | c | ✓ | | | | | | | | | |
| [13] | 2020 | | | ✓ | ✓ | | | | ✓ | | ✓ | ✓ | ✓ | | | ✓ | ✓ | | | | | | | d | ✓ | | | | | | | | | |

Table 3: A classification of the reviewed papers

In what follows, we present a detailed analysis of these classified solutions.

*5.3. Analysis of reviewed MTSP's solutions*

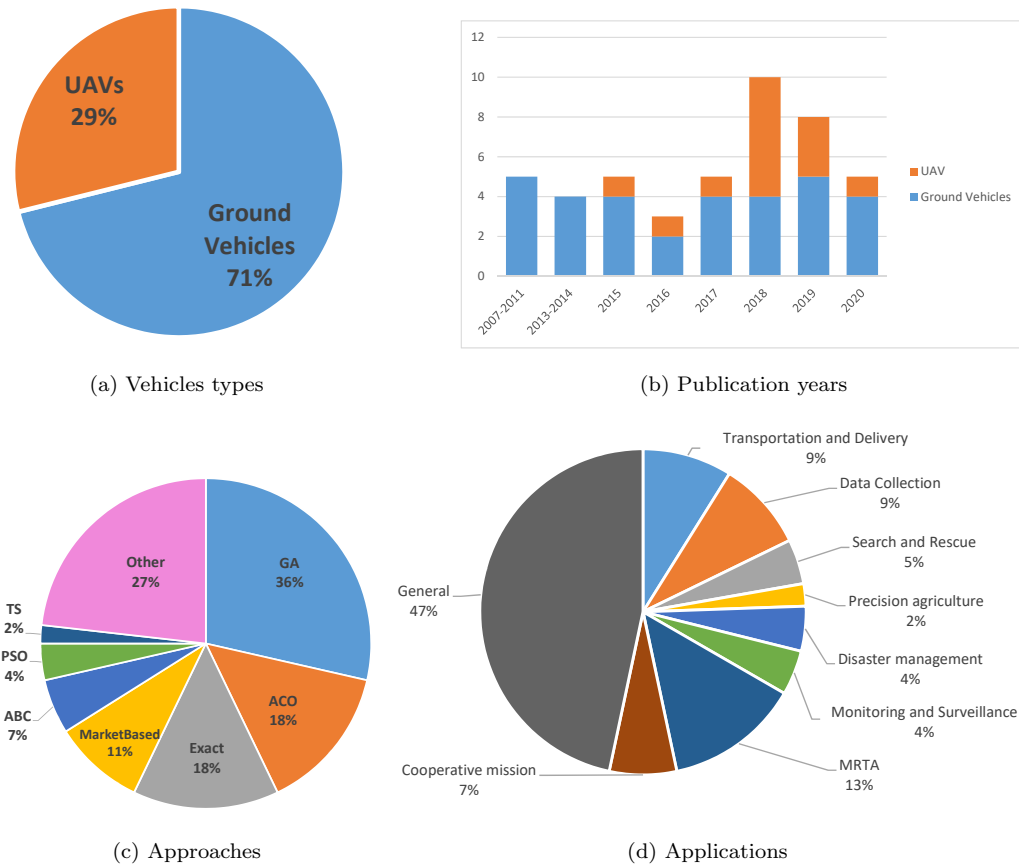

Figure 5: Statistical study of the reviewed papers

In order to better understand and analyse the classification proposed in Table 3, we propose in Figure 5 a statistical study for the papers cited in our survey in terms of the vehicle types, the applied approaches, the application fields, and the publication years.

We can see that 71% of the studies focusing on MTSP are in the context of ground vehicles compared with 29% for flying vehicles as shown in Figure 5a. This can be explained by the fact that MTSP studies for ground vehicles have been published since 2007, however, the oldest paper focusing on UAVs and cited in our survey was published in 2015 as illustrated in Figure 5b. MTSP for ground vehicles and robots is more studied than MTSP for UAVs, since this flying vehicles are considered as an emerging technology and their use in civilian areas is relatively recent.

Moreover, as shown in Figure 5c, the GA approaches is the most used approach in both types of vehicles (36% of papers have used GA). The second most used approaches are exact and ACO approaches (used by 18% of papers). Then, it comes the Market-based approach (used by 11% of papers).

Moreover, as shown in Table 3, for the ground vehicles context, the majority of papers consider the multiple depots variant, however, for the UAVs context the majority of papers consider the single depot variant of MTSP. Indeed, the number of depots used depends on the application domain and also on the type of vehicles. For example, UAVs have a limited autonomy compared with ground vehicles, and these UAVs are generally deployed in a non-extended area. This justifies the need of a single depot to charge and launch these UAVs. Furthermore, this single depot can be mobile in order to optimize both the scheduled time and the energy of the UAVs, for example by using trucks and UAVs for the last mile delivery as proposed in [13, 14].

Moreover, what characterize UAVs context is that some papers such as [32, 33] consider refueling point in the path of UAVs.

Another characteristic of the UAVs solutions is that they consider more constraints, such as the energy constraint in [33, 88, 87, 13] and the time win-dow constraint in [15, 23, 85, 86]. This is justified by the fact that UAVs have more constrained resources (i.e. especially in terms of energy) than ground vehicles. For ground vehicles, the energy is considered as an optimization objective (i.e. to be minimized) rather than a constraint. Moreover, contrary to ground vehicles context, where the single objective as well as the multi-objective variant of MTSP are fairly considered, in the UAVs context the multi-objective variant is rarely considered (only Ref. [26]). Indeed, the most considered objective in UAVs solutions is minimizing the mission time.

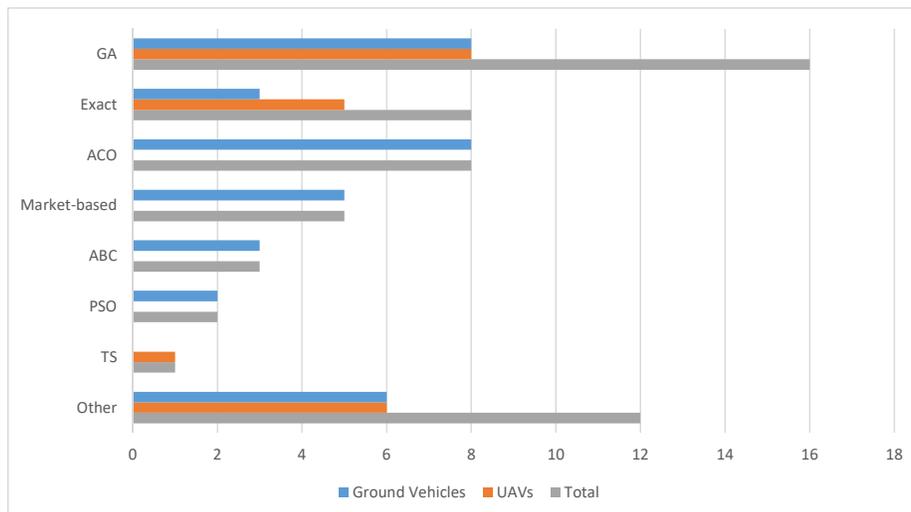

Figure 6: Distribution of solutions per approaches for each vehicle type

Regarding the used approaches, as shown in Figure 6, the genetic algorithm approach is the most used one in both types of vehicles. The second most used approach for ground vehicles is the ACO meta-heuristic. Then, it comes the Market-based approach. It is worth noting that the GA approach was combined with other techniques in several papers, such as GA+ACO in [61, 67], or GA+memetic algorithm in [68]. Moreover, for the UAVs context, the GA approach is widely used with the clustering technique such as in [18, 23, 24, 86, 14]. In addition, we note from Table 3 that MTSP solutions proposed for the UAVs context are more complex and are based on the integration of several approaches. Indeed, UAVs are limited by their energy and load/carrying capacity, and can operate in tandem with trucks, as some papers suggest. The optimization problem therefore becomes complex. In order to optimize the expected time while taking into account the constraints of UAVs and the requirements of the application, many studies have proposed to solve the optimization problem in several steps to reduce the complexity of the problem. To do this, these studies have applied or combined several approaches for these different stages.

In terms of application fields (Figure 5d), the most papers proposed for ground vehicles considered the general context, without being limited to a particular application area, except some papers such as [57, 71, 60, 38, 51], which are proposed for the MRTA problem. However, UAVs solutions present a diversity of applications areas (Table 3).

## 6. Discussion and Future Directions

The previous two sections have been devoted to reviewing contributions proposed for MTSP. Indeed, we have provided an overview of the different approaches proposed in the literature to address MTSP while highlighting the application's areas.

Even though MTSP is very relevant for real-life applications, we pointed out that several studies have solved the general context MTSP without considering a specific application. However, when the study is within a given context such as parcel delivery, data collection, monitoring, and surveillance, etc., new variants of MTSP are proposed. These variants consider different types of vehicles (e.g. robots, vehicles, trucks, and UAVs) along with new characteristics of the depot and the cities to be visited.

For instance, UAVs are characterized by their low cost, high mobility, and can easily be deployed to carry out a search and rescue mission or monitor a given area. However, in the field of transport, ground vehicles and trucks are used because they have great autonomy and load capacity. It should be noted that several studies have been based on heterogeneous vehicles and have proposed solutions where trucks and UAVs are used together to accomplish a given task.

Based on the application requirements, additional constraints are also considered in MTSP, such as vehicle capacity, energy consumption, and time window, which are commonly used in vehicle routing problems. Even if MTSP becomes similar to VRP when considering these constraints, in our survey, we only reviewed papers in which the problem is formulated as MTSP.

In the reviewed studies, several exact and heuristic approaches have been proposed, either to solve MTSP directly or after transformation to TSP.

Although exact approaches give the optimal solution, they are useful only for very small instances due to the NP-hardness of the problem. Metaheuristic approaches, including the genetic algorithm (GA), ant colony optimization (ACO), artificial bee colony (ABC), and particle swarm optimization (PSO), have been extensively explored to efficiently solve MTSP. Nevertheless, the execution time is very long when the problem scales, making it unsuitable for real-time applications. We pointed out that the genetic al-gorithm has been the most widely used meta-heuristic, however, in recent years, there has been a tendency to apply the ACO algorithm.

Besides, several studies proposed to solve MTSP based on hybrid algorithms that combine the use of a meta-heuristic with, for example, a local search or clustering algorithm. The problem is then solved in different phases, so that, the computation complexity is reduced and the convergence time becomes reasonable.

The market-based approach has also been adopted to solve MTSP, as it can deal effectively with dynamic system changes and does not require prior knowledge of all system states to provide a solution.

The path followed by the vehicles to accomplish a given mission must meet two important conditions. On the one hand, the vehicle trajectory must be feasible, hence the need to take into account vehicle constraints such as endurance, kinematics (e.g. speed and acceleration), and system dynamics. On the other hand, the safety of the vehicle must be ensured by avoiding collisions with obstacles as well as collisions between vehicles.

For the sake of simplicity, many studies did not consider these vehicle characteristics and proposed optimization problems that relax these constraints. Only a few contributions on UAV's applications have included the vehicle constraints in the problem model [11] or have considered collision avoidance in the solution [32].

Future studies on the problem of MTSP optimization need to pay more attention to vehicle characteristics and constraints to provide effective solutions for real-life applications.

Unlike vehicles and trucks, ground robots and UAVs might have limited energy. Some studies have focused on UAV's endurance, however, an energy consumption model for robots and UAVs needs to be proposed in future contributions.

As for the classical MTSP, studies on MTSP for UAVs can provide solutions for the data Benchmark [89] to perform comparisons between the different approaches proposed to solve MTSP and to push research studies in that direction.

## 7. Conclusion

The multiple travelling salesman problem is one of the most interesting combinatorial optimization problems due to its ability to describe and formulate real-life applications. Indeed, this survey showed that MTSP is used to formulate optimization problems in several fields, including transportation and delivery, data collection, search and rescue, multi-robot task allocation and scheduling, etc. Although the MTSP importance, there is a lack of a survey that describes existing solutions. This paper aims to fill this gap by providing a comprehensive review of existing and recent solutions for MTSP. We have divided existing solutions into two broad classes: (1) MTSP for vehicles and robots and (2) MTSP for UAVs or drones. Moreover, each class's solutions are classified according to the optimization approaches used, such as exact, meta-heuristic, market-based, etc. This paper also proposed a taxonomy of studied solutions according to several criteria, including MTSP variants, approaches, applications, etc. Finally, it is worth noting that MTSP remains a promising research field, especially for UAVs based applications, where new optimizing problems are emerging.

**The final version of this paper is published in Computer Science Review,** https://doi.org/10.1016/j.cosrev.2021.100369

The final version of this paper is published in Computer Science Review, https://doi.org/10.1016/j.cosrev.2021.100369[25] W. Zhao, Q. Meng, P. W. Chung, A heuristic distributed task allocation method for multivehicle multitask problems and its application to search and rescue scenario, IEEE transactions on cybernetics 46 (4) (2015) 902–915.

[26] S. Hayat, E. Yanmaz, T. X. Brown, C. Bettstetter, Multi-objective UAV path planning for search and rescue, Proceedings - IEEE International Conference on Robotics and Automation (2017) 5569–5574 doi:10.1109/ICRA.2017.7989656.

[27] J. Conesa-Muñoz, G. Pajares, A. Ribeiro, Mix-opt: A new route operator for optimal coverage path planning for a fleet in an agricultural environment, Expert Systems with Applications 54 (2016) 364–378.

[28] J. Conesa-Muñoz, J. M. Bengochea-Guevara, D. Andujar, A. Ribeiro, Route planning for agricultural tasks: A general approach for fleets of autonomous vehicles in site-specific herbicide applications, Computers and Electronics in Agriculture 127 (2016) 204–220.

[29] X. Li, Z. Ma, X. Chu, Y. Liu, A Cloud-Assisted Region Monitoring Strategy of Mobile Robot in Smart Greenhouse, Mobile Information Systems 2019. doi:10.1155/2019/5846232.

[30] O. Cheikhrouhou, A. Koubaa, A. Zarrad, A cloud based disaster management system, Journal of Sensor and Actuator Networks 9 (1) (2020) 6.

[31] S. Trigui, O. Cheikhrouhou, A. Koubaa, A. Zarrad, H. Youssef, An analytical hierarchy process-based approach to solve the multi-objective
66